\documentclass[hyper]{JHEP} % 10pt is ignored!
\usepackage{epsfig}
\usepackage{amsfonts}
\newcommand\fverb{\setbox\pippobox=\hbox\bgroup\verb}
\newcommand\fverbdo{\egroup\medskip\noindent%
            \fbox{\unhbox\pippobox}\ }

\newcommand\fverbit{\egroup\item[\fbox{\unhbox\pippobox}]}
\newbox\pippobox
\title{On Rotating and Oscillating Four-Spin Strings in $AdS_5 \times S^5$}
\author{Kamal L. Panigrahi\\
Department of Physics and Meteorology, \\
Indian Institute of Technology Kharagpur,\\
Kharagpur-721302, INDIA, \\
and  \\
The Abdus Salam International Centre for Theoretical Physics, \\
Strada Costiera 11, Trieste, ITALY \\
E-mail: \email{panigrahi@phy.iitkgp.ernet.in}}
\author{Pabitra M. Pradhan \\
Department of Physics and Meteorology, \\
Indian Institute of Technology Kharagpur,\\
Kharagpur-721 302, INDIA \\
E-mail: \email{ppabitra@phy.iitkgp.ernet.in}} \abstract{We study
rigidly rotating strings in $AdS_5 \times S^5$ background with one
spin along $AdS_5$ and three angular momenta along $S^5$. We find
dispersion relations among various charges and interpret them as
giant magnon and spiky string solutions in various limits. Further
we present an example of oscillating string which oscillates in the
radial direction of the $AdS_5$ and at the same time rotates in
$S^5$.} 

\begin{document}
\section{Introduction}
Recent advances on both the string and the gauge theory sides of
the AdS/CFT correspondence \cite{Maldacena:1997re},
\cite{Gubser:1998bc}, \cite{Witten:1998qj} has added much of our
understanding to the gauge/gravity duality. The duality aims at
establishing the equivalence between the spectrum of anomalous
dimensions of gauge invariant operators in the gauge theory and
the energy spectrum of the string theory states. In case of
$AdS_5 /CFT_4$ duality, it is possible to see how certain simple string states
actually appear as field theory operators \cite{Berenstein:2002jq}
\cite{Gubser:2002tv}. This
state/operator duality becomes tractable in the strong coupling limit
or the so called semiclassical limit. In this limit the appearance of
integrable structures on both sides of
the duality makes it simple to study and to make further
predictions regarding the correspondence. The integrability arises
as a quantum symmetry of operator mixing in CFT side \cite{Minahan:2002ve}, \cite{Beisert:2003yb}
and as a classical symmetry on the string world-sheet in AdS space\cite{Bena:2003wd}. More
precisely, the integrability has improved the understanding of the
equivalence between the Bethe equation for the spin chain and the
corresponding classical realization of Bethe equation for the
classical $AdS_5 \times S^5$ string sigma model, see e.g.
\cite{Kazakov:2004qf},
\cite{Zarembo:2004hp}. Study of the
multi-spin rotating string solutions, e.g. in $AdS_5 \times S^5$
\cite{Frolov:2002av}, \cite{Frolov:2003qc}, \cite{Frolov:2003tu}, \cite{Tseytlin:2003ii} and the Bethe
equation for the diagonalization of the integrable spin chain in
the SYM theory, e.g. \cite{Minahan:2002ve},
\cite{Beisert:2003yb},
\cite{Beisert:2004ry}
have been studied in great detail.
In the semiclassical limit, the study of AdS/CFT duality has triggered
much of interest in recent past. In this connection, magnons,
which are elementary excitations on the spin chain have been
realized in \cite{Hofman:2006xt} as dual to specific rotating
semiclassical string states on $R_t \times S^2$. Soon both
infinite and finite long spin chain \cite{Tseytlin:2003ii},
\cite{Tseytlin:2004xa},
\cite{Ishizeki:2007we}
operators have been mapped to different string states in various
backgrounds and magnon bound states \cite{Arutyunov:2006gs},
\cite{Minahan:2006bd}, \cite{Dorey:2006dq}, \cite{Chen:2006gea},
\cite{Kruczenski:2006pk}, \cite{Ryang:2006yq} dual to strings on different subspaces of
$AdS_5 \times S^5$ with two and three angular momenta has been
found out. Infact a general class of rotating string solutions
known as spiky string was found out in
\cite{Kruczenski:2004wg} which
correspond to a higher twist operator from the field theory view
point. Further it was also observed in \cite{Ishizeki:2007we} that both
giant magnon and single spike solutions can be viewed as two
different limits of the same rigidly rotating strings on $S^2$ and
$S^3$. In an attempt to find out new solution, a large class of multispin spiky string
and giant magnon solutions have been studied, in various
backgrounds including the less supersymmetric, orbifolded and non-AdS backgrounds, for
example in
\cite{Lunin:2005jy}, \cite{Bobev:2005cz}, \cite{Chu:2006ae}, \cite{Bobev:2006fg}, \cite{Kluson:2007qu}, \cite{Lee:2008sk}, \cite{David:2008yk}, \cite{Grignani:2008is}, \cite{Lee:2008ui}, \cite{Ryang:2008rc}, \cite{Benvenuti:2008bd}, \cite{Abbott:2008qd}, \cite{Biswas:2011wu}, \cite{Biswas:2012wu}.
More recently in \cite{Giardino:2011dz}, \cite{Panigrahi:2011be},
\cite{Giardino:2011uc}, \cite{Panigrahi:2012bm} more general three
spin giant magnons and spiky strings have been studied and interesting
dispersion relation among the divergent energy and angular momenta have
been obtained. In particular in
\cite{Giardino:2011uc} an interesting rotating solution has been studied
with thee divergent angular momenta along the S$^5$ and divergent
deficit angles around the sphere in $AdS_5 \times S^5$. Here, we would like to
generalize the result to add another spin along AdS$_5$. We also wish to
generalize the four spin solutions to a class of giant magnon obtained in
\cite{Kruczenski:2006pk} and show the similarity of the
dispersion relation to the solutions of \cite{Ryang:2006yq} with three spin
giant magnon. We wish to note that the solutions presented here
correspond to rotating open strings with Neumann boundary condition.

We also describe a class of multispin oscillating string in $AdS_5
\times S^5$. The oscillating string solutions found in \cite{Gubser:2002tv}
describes a string oscillating in one plane. In \cite{Minahan:2002rc} and
\cite{Beccaria:2010zn}, pulsating string solutions in $AdS_5$
and $S^5$ have been worked out separately where as in
\cite{Park:2005kt}, rotating and oscillating strings in $AdS_5$
have been derived. In these papers, the solutions to the equations of
motion for a general rotating and oscillating string have been discussed and
the energy expression as the function of oscillation number has
been derived. We wish to generalize the solutions, to describe
the motion of a string which oscillates in the $S^5$ with a
constant $\rho$ value in $AdS_5$, in subsequent section.

The rest of the paper is organized as follows. In section-2 we
solve the general equation of motion for a rigidly rotating string
in the background of $AdS_5 \times S^5$, with one spin along
AdS$_5$ and three angular momenta along $S^5$ directions. We write
down the most general solutions to the equations of motion of the
fundamental string and write down the conserved quantities. In
section-3 we compute all the charges and have shown a class of
spiky string solutions with the help of a set of integration
constants. In section-4 we present a different class of solutions
with a different set of integration constants. In section-5, we
present an oscillating string which oscillates in the $S^5$.
Finally in section-6, we conclude with some discussion and
remarks.

\section{Rotating Strings with Four Spin in $AdS_5 \times S^5$}
The full metric for $AdS_5 \times S^5$ background is
\begin{eqnarray}
ds^2 &=& -\cosh^2\rho dt^2 + d\rho^2 + \sinh^2\rho (d\varphi^2_3 +
\cos^2\varphi_3 d\varphi^2_2 + \sin^2 \varphi_3 d\varphi^2_1) +
d\psi^2 \cr & \cr && + \sin^2\psi d\theta^2 +\sin^2\psi
\cos^2\theta d\phi_1^2 +\cos^2\psi d\phi_2^2 + \sin^2\psi
\sin^2\theta d\phi_3^2, \label{1}
\end{eqnarray}
where $\rho \in [0,\infty],~~\theta,\psi \in [0, \frac{\pi}{2}]$,
 $\phi_1, \phi_2, \phi_3 \in [0,2\pi]$ and $\varphi_1, \varphi_2,
 \varphi_3$ are co-ordinates on a three sphere. For studying the
 rotating string with three angular momenta along S$^5$ and one
 spin along AdS$_5$,
%We get the relevant
%part of the $ AdS_5 \times S^5$ metric
%for the rotating strings with three angular momenta along S$^5$
%and one spin along AdS$_5$ by
we substitute $\varphi_1 = \varphi_2
= 0, \varphi_3 = \varphi$ and $\psi$ as a constant. The Polyakov
action for the fundamental string in the background (\ref{1})
with the above identification is given by
\begin{eqnarray}
I &=& \frac{T}{2}\int d\tau d\sigma\Big[-\cosh^2\rho (\dot{t}^2 -
{t^{\prime}}^2) + \dot{\rho}^2 - {\rho^{\prime}}^2 +
\sinh^2\rho(\dot{\varphi}^2 - {\varphi^{\prime}}^2) +\sin^2\psi (
\dot{\theta}^2 - {\theta^{\prime}}^2) \cr & \cr && +
\sin^2\psi\cos^2\theta(\dot{\phi_1}^2 - {\phi_1^{\prime}}^2) +
\cos^2\psi(\dot{\phi_2}^2 - {\phi_2^{\prime}}^2)  +
\sin^2\psi\sin^2\theta(\dot{\phi_3}-\phi_3^{\prime}\Big],
\label{2}
\end{eqnarray}
where the dot and prime denote the derivatives with respect to
$\tau$ and $\sigma$ respectively and the string tension $T$ is a
function of 't Hooft coupling $\lambda$ as $ T =
\frac{\sqrt{\lambda}}{2\pi}$. We choose the following ansatz for
the rotating string\footnote{we describe the open string
solutions with Neumann boundary conditions.}
\begin{eqnarray}
&&t = \tau + f_1(y), ~~~~ \rho = \rho(y), ~~~~ \psi \in
\Big(0,\frac{\pi}{2}\Big),~~~~\theta = \theta(y),~~~~  \varphi
=\omega\Big(\tau + f_2(y)\Big),\cr & \cr && \phi_1 = \omega_1\tau
+ g_1(y), ~~~~~~~\phi_2 = \omega_2\tau + g_2(y),~~~~~~~ \phi_3 =
\omega_3\tau + g_3(y), \label{3}
\end{eqnarray}
where $y$ is a function of world sheet coordinates, $y = a\sigma
- b\tau$.\\
From the equations of motion for $t,~\varphi,~\phi_1,~\phi_2$ and
$\phi_3$, we get the expressions for $f_1,~f_2,~g_1,~g_2$ and
$g_3$ which are
\begin{eqnarray}
f_{1y} &=& \frac{1}{a^2 - b^2}\left(\frac{A_1}{\cosh^2\rho} -
b\right), \cr & \cr f_{2y} &=& \frac{1}{a^2 -
b^2}\left(\frac{A_2}{\sinh^2\rho} - b\right), \cr & \cr g_{1y} &=&
\frac{1}{a^2 - b^2}\left(\frac{B_1}{\sin^2\psi\cos^2\theta} -
b\omega_1\right), \cr & \cr g_{2y} &=& \frac{1}{a^2 -
b^2}\left(\frac{B_2}{\cos^2\psi} - b\omega_2\right),\cr & \cr
g_{3y} &=& \frac{1}{a^2 -
b^2}\left(\frac{B_3}{\sin^2\psi\sin^2\theta} - b\omega_3\right),
\label{4}
\end{eqnarray}
where $f_y = \frac{\partial f}{\partial y}$, $g_y = \frac{\partial g}{\partial y}$
%denote partial derivatives of $f$ with respect to $y$
and $A_1,~A_2,~B_1,~B_2$ and $B_3$ are integration constants.
Using above (\ref{4}) values we can get the equation of motion for
$\rho$ and $\theta$ which are
\begin{eqnarray}
\rho_{yy} &=& - \frac{1}{(a^2 - b^2)^2}\cosh\rho\sinh\rho
\Big(\frac{A_1^2}{\cosh^4\rho} - \frac{A_2^2\omega^2}{\sinh^4\rho}
- a^2 + a^2\omega^2 \Big), \cr & \cr \theta_{yy} &=& \frac{1}{(a^2
- b^2)^2}\cos\theta\sin\theta
\Big(\frac{B_3^2}{\sin^4\psi\sin^4\theta} -
\frac{B_1^2}{\sin^4\psi\cos^4\theta} + a^2\omega_1^2 -
a^2\omega_3^2 \Big) , \label{5}
\end{eqnarray}
where $\rho_{yy} = \frac{\partial^2 \rho}{\partial y^2}$ etc.
Now, the first Virasoro constraint $T_{\tau\sigma} = 0$ gives
\begin{eqnarray}
\rho_y^2 + \sin^2\psi \theta_y^2 &=& \cosh^2\rho(
f_{1y}^2-\frac{1}{b}f_{1y}) - \omega^2
\sinh^2\rho(f_{2y}^2-\frac{1}{b}f_{2y}) \cr & \cr &&-
\sin^2\psi\cos^2\theta( g_{1y}^2-\frac{\omega_1}{b}g_{1y}) -
\cos^2\psi( g_{2y}^2-\frac{\omega_2}{b}g_{2y})\cr & \cr
&&-\sin^2\psi\sin^2\theta(g_{3y}^2-\frac{\omega_3}{b}g_{3y}),
\label{6}
\end{eqnarray}
and the second virasoro constraint $T_{\tau\tau} +
T_{\sigma\sigma} = 0$ gives
\begin{eqnarray}
\rho_y^2 + \sin^2\psi \theta_y^2 &=& \cosh^2\rho\Big(
f_{1y}^2+\frac{1-2bf_{1y}}{a^2+b^2}\Big) - \omega^2
\sinh^2\rho\Big(f_{2y}^2+\frac{1-2bf_{2y}}{a^2+b^2}\Big) \cr & \cr
&&- \sin^2\psi\cos^2\theta\Big(
g_{1y}^2+\frac{\omega_1^2-2b\omega_1 g_{1y}}{a^2+b^2}\Big)-
\cos^2\psi\Big( g_{2y}^2-\frac{\omega_2^2-2b\omega_2
g_{2y}}{a^2+b^2}\Big)\cr & \cr &&-\sin^2\psi\sin^2\theta\Big(
g_{3y}^2+\frac{\omega_3^2-2b\omega_3 g_{3y}}{a^2+b^2}\Big).
\label{7}
\end{eqnarray}
The conserved quantities are
\begin{eqnarray}
E &=& T \int d\sigma \cosh^2\rho ~\dot{t}, \cr && \cr S &=& T \int
d\sigma \sinh^2\rho ~\dot{\varphi}, \cr && \cr \jmath_1 &=& T \int
d\sigma \sin^2\psi \cos^2\theta ~\dot{\phi_1}, \cr && \cr \jmath_2
&=& T \int d\sigma \cos^2\psi ~\dot{\phi_2}, \cr && \cr \jmath_3
&=& T \int d\sigma \sin^2\psi \sin^2\theta ~\dot{\phi_3}.
\label{8}
\end{eqnarray}
Deficit angle is defined as $ \Delta\phi = \int \frac{\partial
\phi}{\partial y} dy. $ Now, we define $S_1 = \frac{S}{\omega}$,
  $J_1 = \frac{\jmath_1}{\sin^2\psi}$,  $ J_2 = \frac{\jmath_2}{\cos\psi
  \sin\psi}$, $J_3 = \frac{\jmath_3}{\sin^2\psi}$ and
  $\Delta\phi_2 = \frac{\Delta\phi_2}{\tan\psi}$.
%These definitions have used in the section-3 and section-4.
From the Virasoro constraints
(\ref{6}) and (\ref{7}), we get the following relation among the integration
constants
\begin{equation}
A_1-A_2\omega^2-B_1\omega_1 -  B_2\omega_2 - B_3\omega_3 = 0.
\label{9}
\end{equation}
For the convenience of our solution, the arbitrary parameters
$A_1$ and $A_2$ that characterize the time and angle coordinates
of string in $AdS_3$ are chosen as $A_1 = b$ and $A_2 = 0.$ Using
this in the equation of motion of $\rho$ (\ref{5}) and integrating
it once with a zero integration constant we get
\begin{equation}
\rho^2_y = \frac{1}{(a^2-b^2)^2}
\sinh^2\rho\Big[a^2-a^2\omega^2-\frac{b^2}{\cosh^2\rho}\Big].
\label{10}
\end{equation}
Subtracting the above equation (\ref{10}) from the second Virasoro
constraint (\ref{7}), we get
\begin{equation}
{\theta_y}^2 +
\frac{c^2_1}{\cos^2\theta}+c^2_2+\frac{c^2_3}{\sin^2\theta}+\upsilon^2_1\cos^2\theta+\upsilon^2_2
+ \upsilon^2_3\sin^2\theta - \kappa^2 = 0, \label{11}
\end{equation}
where
\begin{eqnarray}
c_1 &=& \frac{B_1}{(a^2-b^2)\sin^2\psi},~~~   c_2 =
\frac{B_2}{(a^2-b^2)\sin\psi\cos\psi},~~~ c_3 =
\frac{B_3}{(a^2-b^2)\sin^2\psi}, \cr && \cr \upsilon_1 &=&
\frac{a\omega_1}{a^2-b^2},~~~~~~~~~~~~~ \upsilon_2 =
\frac{a\omega_2}{a^2-b^2}\frac{\cos\psi}{\sin\psi},~~~~~~~~~~~~
\upsilon_3 = \frac{a\omega_3}{a^2-b^2}, \cr && \cr \kappa^2 &=&
\frac{a^2+b^2}{(a^2-b^2)^2}\frac{1}{\sin^2\psi}. \label{12}
\end{eqnarray}
For concrete realization of the rotating string solutions presented above as the
giant magnon and single spike solutinos, in what follows, we wish to study two different
sets of integration constants values as follows
\begin{equation}
{\rm Case~ I :}~B_1 \neq B_2 \neq B_3 \neq 0, \label{13}
\end{equation}
\begin{equation}
{\rm Case ~II:}~B_1 = B_2 = 0.\label{14}
\end{equation}
\section{Case I}
Now, we substitute $\xi = \cos2\theta$ in the equation (\ref{11})
and rewrite it
\begin{eqnarray}
\xi^2_y &=& 2(\upsilon^2_1-\upsilon^2_3)\xi^3 +
2(\upsilon^2_1+\upsilon^2_3-2m)\xi^2 + 2\Big(\upsilon^2_3 -
\upsilon^2_1 +4(c^2_1-c^2_3)\Big)\xi \cr && \cr &&-2(\upsilon^2_1
+ \upsilon^2_3) - 8(c^2_1 + c^2_3) + 4m, \label{15}
\end{eqnarray}
where $m = \kappa^2 -\upsilon^2_2 - c^2_2$. In order to get the
solution, we choose $c^2_1 = \frac{1}{4}(m-\upsilon^2_1)$ and
$c^2_3 = \frac{1}{4}(m-\upsilon^2_3)$. Using these values in
(\ref{15}), we get
\begin{equation}
dy =
\frac{1}{\sqrt{2(\upsilon^2_3-\upsilon^2_2)}}\frac{d\xi}{\xi\sqrt{\xi_0
- \xi}},
  \label{16}
\end{equation}
where $\xi_0 = \frac{\upsilon^2_1 + \upsilon^2_3 -
2m}{\upsilon^2_3 - \upsilon^2_1}$ and $\xi_0 \in (0,1)$. Now, the
conserved quantities including that of the angle differences between the
end point of the string can be rewritten as
\begin{eqnarray}
E &=& \frac{T}{a}\int dy  + \frac{aT}{a^2-b^2} \int dy~
\sinh^2\rho, \cr && \cr S_1 &=& \frac{aT}{a^2-b^2} \int dy~
\sinh^2\rho, \cr && \cr J_1 &=& (\frac{1}{2}a\upsilon_1 -
bc_1)(E-S_1) +
\frac{1}{2}\frac{T\upsilon_1}{\sqrt{2(\upsilon^2_3-\upsilon^2_1)}}
\int_0^{\xi_0} \frac{d\xi}{\sqrt{\xi_0-\xi}}, \cr && \cr J_2 &=&
(a\upsilon_2 -bc_2)(E-S_1), \cr && \cr J_3 &=& (\frac{1}{2}
a\upsilon_3-bc_3)(E-S_1) - \frac{1}{2}
\frac{T\upsilon_3}{\sqrt{2(\upsilon^2_3-\upsilon^2_1)}}\int_0^{\xi_0}
\frac{d\xi}{\sqrt{\xi_0-\xi}}, \cr && \cr T\Delta\phi_1 &=& (2ac_1
- b\upsilon_1)(E-S_1) -
\frac{2c_1T}{\sqrt{2(\upsilon^2_3-\upsilon^2_1)}} \int_0^{\xi_0}
\frac{1}{\xi + 1}\frac{d\xi}{\sqrt{\xi_0-\xi}}, \cr && \cr
T\Delta\phi_2 &=& (ac_2 - b\upsilon_2)(E-S_1), \cr && \cr
T\Delta\phi_3 &=& (2ac_3 - b\upsilon_3)(E-S_1) -
\frac{2c_3T}{\sqrt{2(\upsilon^2_3-\upsilon^2_1)}} \int_0^{\xi_0}
\frac{1}{\xi - 1}\frac{d\xi}{\sqrt{\xi_0-\xi}}. \label{17}
\end{eqnarray}
\subsection{Giant Magnon Solutions}
Here we choose constants values in such a way that we get finite deficit angles
and large energy and angular momenta. Thus
\begin{equation}
2a(c_1+c_3) = b(\upsilon_1 + \upsilon_3),~~~ ac_2 =
b\upsilon_2,~~~ ac_1 \neq b\upsilon_1,~~~ ac_3 \neq b\upsilon_3.
\label{18}
\end{equation}
We take $ J = J_1 + J_3$ and $\Delta\phi = \Delta\phi_1 +
\Delta\phi_3$, so
\begin{eqnarray}
J &=& \frac{1}{2} a(\upsilon_1 + \upsilon_3)(1-
\frac{b^2}{a^2})(E-S_1) + \frac{1}{2}\frac{T(\upsilon_1 -
\upsilon_3)}{\sqrt{2(\upsilon^2_3-\upsilon^2_1)}} \int_0^{\xi_0}
\frac{d\xi}{\sqrt{\xi_0-\xi}}, \cr && \cr J_2 &=&
a\upsilon_2(1-\frac{b^2}{a^2})(E-S_1), ~~~~~~~ \Delta\phi_2 = 0
,\cr && \cr \Delta\phi &=& \sqrt{\frac{2}{\upsilon^2_3 -
\upsilon^2_1}} \int_0^{\xi_0}\frac{d\xi}{\sqrt{\xi_0 -
\xi}}(\frac{c_3}{1-\xi}-\frac{c_1}{1+\xi}). \label{19}
\end{eqnarray}
With the constraint $ 4\upsilon^2_2 + (\upsilon_1 + \upsilon_3)^2
= \frac{4a^2}{(a^2 - b^2)^2}$, we get the magnon dispersion
relation
\begin{equation}
\sqrt{(E-S_1)^2 - J^2_2} -J = T\sin\Delta\phi, \label{20}
\end{equation}
where deficit angle has a constraint $\sin\Delta\phi =
\sqrt{\frac{\upsilon_3 - \upsilon_1}{\upsilon_3 +
\upsilon_1}}\sqrt{\frac{\xi_0}{2}}$.
\subsection{Spiky String Solutions}
Similarly, to get the finite momenta and large energy and deficit
angles, we choose the conditions as
\begin{equation}
2b(c_1 + c_3) = a(\upsilon_1 + \upsilon_3),~~~ bc_2 =
a\upsilon_2,~~~ a\upsilon_1 \neq bc_1,~~~ a\upsilon_3 \neq bc_3.
\label{21}
\end{equation}
So
\begin{eqnarray}
J &=& \frac{1}{2}\frac{T(\upsilon_1 -
\upsilon_3)}{\sqrt{2(\upsilon^2_3-\upsilon^2_1)}} \int_0^{\xi_0}
\frac{d\xi}{\sqrt{\xi_0-\xi}}, \cr && \cr T\Delta\phi &=&
(\frac{a^2}{b^2}-1)b(\upsilon_1 +\upsilon_3)(E - S_1) -
\frac{2T}{\sqrt{2(\upsilon^2_3 - \upsilon^2_1)}}
\int_0^{\xi_0}\frac{d\xi}{\sqrt{\xi_0 -
\xi}}(\frac{c_1}{\xi+1}-\frac{c_3}{\xi-1}), \cr && \cr
T\Delta\phi_2 &=& (\frac{a^2}{b^2}-1)b\upsilon_2(E-S_1),~~~~~~ J_2
= 0.  \label{22}
\end{eqnarray}
With the constraint $ \upsilon^2_2 + (\upsilon_1 + \upsilon_3)^2 =
\frac{b^2}{(a^2 - b^2)^2}$, we get the spiky dispersion relation
\begin{equation}
\sqrt{(E-S_1)^2 - (T\Delta\phi_2)^2} - T\Delta\phi =
2T(\frac{\pi}{2} - \theta_0), \label{23}
\end{equation}
where $$ \theta_0 = \frac{\pi}{2} -\frac{1}{\sqrt{2(\upsilon^2_3 -
\upsilon^2_1)}} \int_0^{\xi_0} \frac{d\xi}{\sqrt{\xi_0 -
\xi}}\left(\frac{c_1}{\xi+1}-\frac{c_3}{\xi-1}\right), ~~~{\rm and}~~ \xi_0 =
\cos2\theta_0 \ .$$
\section{Case II}
Putting condition(\ref{14}) in the equation (\ref{11}), we get
\begin{equation}
\theta^2_y = \kappa^2_1  - c^2_3\frac{\cos^2\theta}{\sin^2\theta}
- (\upsilon^2_1 + \upsilon^2_2) - (\upsilon^2_3- \upsilon^2_1)
\sin^2\theta,  \label{24}
\end{equation}
where $ \kappa^2_1 = \frac{(a^2+b^2)\omega^2_3 \sin^2\psi -
b^2}{(a^2 -b^2)^2\omega^2_3\sin^4\psi}$ and we choose $ \kappa^2_1
= \upsilon^2_2 + \upsilon^2_3$. Now, equation(\ref{24}) can be
written as
\begin{equation}
\theta^2_y = \frac{\upsilon^2_3 -
\upsilon^2_1}{\sin^2\theta}(\sin^2\theta -
\sin^2\theta_1)(\sin^2\theta_0 - \sin^2\theta), \label{25}
\end{equation}
where
\begin{equation}
\sin^2\theta_0 = \frac{c^2_3}{\upsilon^2_3 - \upsilon^2_1}, ~~~~~
\sin^2\theta_1 = \frac{\kappa^2_1 - \upsilon^2_1 -
\upsilon^2_2}{\upsilon^2_3 - \upsilon^2_1}.   \label{26}
\end{equation}
From (\ref{26}), We  get the two conditions (i) $\kappa^2_1 =
\upsilon^2_2 + \upsilon^2_3$ and (ii) $c^2_3 = \upsilon^2_3 -
\upsilon^2_1$. Both the conditions will give us same $dy$
equation, only with different $\theta_0$ value. So the conserved
quantities will remain same and this will lead us to same
relations. Here we study the different dispersion relations with
the (i) condition only. So taking $\kappa^2_1 = \upsilon^2_2 +
\upsilon^2_3$, the equation (\ref{25}) changes to
\begin{equation}
dy = \frac{1}{\sqrt{\upsilon^2_3 - \upsilon^2_1}}
\frac{\sin\theta}{\cos\theta} \frac{d\theta}{\sqrt{\sin^2\theta -
\sin^2\theta_0}}  .\label{27}
\end{equation}
Thus, the conserved quantities are rewritten as
\begin{eqnarray}
E &=& \frac{T}{a}\int dy  + \frac{aT}{a^2-b^2} \int dy~
\sinh^2\rho, \cr && \cr S_1 &=& \frac{aT}{a^2-b^2} \int dy~
\sinh^2\rho, \cr && \cr J_1 &=&
\frac{T\upsilon_1}{\sqrt{\upsilon^2_3 - \upsilon^2_1}}
\int_{\theta_0}^{\frac{\pi}{2}} \frac{\sin\theta \cos\theta
d\theta}{\sqrt{\sin^2\theta - \sin^2\theta_0}}, \cr && \cr J_2 &=&
a\upsilon_2(E -S_1), \cr && \cr J_3 &=& (a\upsilon_3 - bc_3) (E -
S_1) - \frac{T\upsilon_3}{\sqrt{\upsilon^2_3 - \upsilon^2_1}}
\int_{\theta_0}^{\frac{\pi}{2}} \frac{\sin\theta \cos\theta
d\theta}{\sqrt{\sin^2\theta - \sin^2\theta_0}}, \cr && \cr
T\Delta\phi_1 &=& -b\upsilon_1 (E - S_1), \cr && \cr T\Delta\phi_2
&=& -b\upsilon_2 (E - S_1), \cr && \cr T\Delta\phi_3 &=& (ac_3 -
b\upsilon_3)(E -S_1) + \frac{Tc_3}{\sqrt{\upsilon^2_3 -
\upsilon^2_1}} \int_{\theta_0}^{\frac{\pi}{2}}
\frac{\cos\theta}{\sin\theta} \frac{ d\theta}{\sqrt{\sin^2\theta -
\sin^2\theta_0}}.  \label{28}
\end{eqnarray}
\subsection{Giant Magnon and Spiky String solutions}
For the region of $~0 \leq 1-\omega^2 \leq \frac{b^2}{a^2}$, we
can rewrite the eqn (\ref{10}) as
\begin{equation}
\rho_y = \pm \frac{\sqrt{1-\omega^2}}{a^2-b^2} a \tanh\rho
\sqrt{\sinh^2\rho-\alpha^2}, \label{A}
\end{equation}
where $\alpha =
\sqrt{\frac{\frac{b^2}{a^2}+\omega^2-1}{1-\omega^2}}$. The
solution for the (\ref{A}) is
\begin{equation}
\sinh\rho = \frac{\alpha}{\cos\beta y}, ~~~~~~~~~~~~~~~~~~~{\rm for} ~~
-\frac{\pi}{2\beta} \leq y \leq \frac{\pi}{2\beta}, \label{B}
\end{equation}
where $\beta = \frac{\sqrt{b^2+a^2(\omega^2-1)}}{a^2-b^2}$.
Similarly for the region of $~\frac{b^2}{a^2} \leq 1-\omega^2$, we
get the solution as
\begin{eqnarray}
\sinh\rho &=& \frac{\alpha}{\sinh\beta y}, ~~~~~~~~~~ {\rm for}~~ 0 \leq
y < \infty, \cr && \cr &=& -\frac{\alpha}{\sinh\beta y},
~~~~~~~~~~{\rm for} ~~-\infty <y \leq 0, \label{C}
\end{eqnarray}
where $\alpha =
\sqrt{\frac{1-\omega^2-\frac{b^2}{a^2}}{1-\omega^2}}$ and $\beta =
\frac{\sqrt{a^2(\omega^2-1) - b^2}}{a^2-b^2}$. Using (\ref{A}),
(\ref{B}) and (\ref{C}), we can compute the deficit angle for time
as
\begin{equation}
\Delta t = -2\arctan \frac{\sqrt{1-\alpha^2}}{\alpha}. \label{D}
\end{equation}
Here $\alpha$ is characterized by a time difference between the
two endpoints of the open string. In order to get the dispersion
relation, we choose the condition among following constants as
\begin{equation}
\upsilon_3 = \frac{1}{2a}(1+bc_3). \label{29}
\end{equation}
Using (\ref{28}) and (\ref{29}), we get the giant magnon
dispersion relation as
\begin{equation}
E - J_3 = S_1 + \frac{\upsilon_3}{\upsilon_1}J_1 +
\frac{\upsilon_3}{\upsilon_2}J_2.   \label{30}
\end{equation}
We can define $J = J_3 + \frac{\upsilon_3}{\upsilon_1}J_1$ and
make the $\Delta\phi_3$ finite by choosing $ac_3 = b\upsilon_3$.
From \cite{Ryang:2006yq}, we can regularize $S_1$ and from
\cite{Arutyunov:2006gs} and \cite{Minahan:2006bd}, we can write
down the finite terms for $\frac{\upsilon_3}{\upsilon_2}J_2$,
which will give us the dispersion relation as
\begin{eqnarray}
(E - J)_{reg} =  - \sqrt{S_{reg}^2 +
\frac{\lambda}{\pi^2}\cos^2\frac{\Delta t}{2}} + \sqrt{J_2^2 +
\frac{\lambda}{\pi^2}\sin^2\frac{\Delta \phi_3}{2}}. \label{31}
\end{eqnarray}
Similarly, we choose the condition for spiky string as
\begin{equation}
\upsilon^2_1 + \upsilon^2_2 + \upsilon^2_3 = \frac{1}{b^2}(1-
a^2c^2_3) + 2 \frac{a}{b}c_3\upsilon_3. \label{32}
\end{equation}
Using (\ref{32}) and (\ref{28}), we get the spiky string
dispersion relation
\begin{equation}
\sqrt{(E - S_1)^2 - (T\Delta\phi_1)^2 - (T\Delta\phi_2)^2} -
T\Delta\phi_3 = 2T(\frac{\pi}{2} - \theta_0).   \label{33}
\end{equation}
\section{Oscillating in S$^5$ with $\rho$ fixed in $AdS$}
Here we consider a string oscillating in the $\psi$ direction of
the $S^5$, which is periodic in nature and $\rho$ is fixed to
$\rho_0$ in $AdS$. We wish to study a configuration with two spin
in $AdS$ and two angular momenta along $S^5$. For the study of
this oscillating string we substitute $\varphi_3 = \frac{\pi}{4}$
and $\phi_1 = \phi_3$ in (\ref{1}) and get the metric as
\begin{eqnarray}
ds^2 &=& -\cosh^2\rho_0 dt^2 + \frac{1}{2} \sinh^2\rho_0 (
d\varphi^2_1 + d\varphi^2_2) +d\psi^2 \cr && \cr &&+ \sin^2\psi
d\theta^2 +\sin^2\psi d\phi_1^2 +\cos^2\psi d\phi_2^2.\label{M}
\end{eqnarray}
Now we parameterize the coordinates as
\begin{equation}
t = \kappa\tau,~~~ \varphi_i = \mu_i\tau,~~~\psi =
\psi(\tau),~~~\theta = m \sigma,~~~ \phi_i = \nu_i\tau, \label{N}
\end{equation}
where $i = 1,2$. The Nambu-Goto action for the string in the
background (\ref{M}) with the above ansatz (\ref{N}) is given
by
\begin{equation}
I = m \sqrt{\lambda} \int ~d\tau ~\sin\psi
\sqrt{\kappa^2\cosh^2\rho_0 - \frac{1}{2} (\mu^2_1 + \mu^2_2)
\sinh^2\rho_0 - \dot{\psi^2} - \nu^2_1 \sin^2\psi - \nu^2_2
\cos^2\psi}.\label{O}
\end{equation}
The conserved quantities derived from the above action (\ref{O}) are
\begin{eqnarray}
E &=& \frac{m\sqrt{\lambda}\kappa \sin\psi
\cosh^2\rho_0}{\sqrt{\kappa^2\cosh^2\rho_0 - \frac{1}{2} (\mu^2_1
+ \mu^2_2) \sinh^2\rho_0 - \dot{\psi^2} - \nu^2_1 \sin^2\psi -
\nu^2_2 \cos^2\psi}}, \cr && \cr S_1 &=&
\frac{m\sqrt{\lambda}\mu_1 \sin\psi
\sinh^2\rho_0}{2\sqrt{\kappa^2\cosh^2\rho_0 - \frac{1}{2} (\mu^2_1
+ \mu^2_2) \sinh^2\rho_0 - \dot{\psi^2} - \nu^2_1 \sin^2\psi -
\nu^2_2 \cos^2\psi}}, \cr && \cr S_2 &=&
\frac{m\sqrt{\lambda}\mu_2 \sin\psi
\sinh^2\rho_0}{2\sqrt{\kappa^2\cosh^2\rho_0 - \frac{1}{2} (\mu^2_1
+ \mu^2_2) \sinh^2\rho_0 - \dot{\psi^2} - \nu^2_1 \sin^2\psi -
\nu^2_2 \cos^2\psi}}, \cr && \cr J_1 &=&
\frac{m\sqrt{\lambda}\nu_1 \sin^3\psi}{\sqrt{\kappa^2\cosh^2\rho_0
- \frac{1}{2} (\mu^2_1 + \mu^2_2) \sinh^2\rho_0 - \dot{\psi^2} -
\nu^2_1 \sin^2\psi - \nu^2_2 \cos^2\psi}}, \cr && \cr J_2 &=&
\frac{m\sqrt{\lambda}\nu_2 \sin\psi
\cos^2\psi}{\sqrt{\kappa^2\cosh^2\rho_0 - \frac{1}{2} (\mu^2_1 +
\mu^2_2) \sinh^2\rho_0 - \dot{\psi^2} - \nu^2_1 \sin^2\psi -
\nu^2_2 \cos^2\psi}}. \label{P}
\end{eqnarray}
From the above equation (\ref{P}), we get a relation among the
conserved charges
\begin{equation}
\frac{E}{\kappa} - \frac{S_1}{\mu_1} -\frac{S_2}{\mu_2} =
\frac{J_1}{\nu_1} + \frac{J_2}{\nu_2}. \label{Q}
\end{equation}
This (\ref{Q}) is the dispersion relation for a string which is
oscillating in the $S^5$ from a minimum value of $\psi_{min}$ to a
maximum $(\frac{\pi}{2} - \psi_{min})$ and at the same time it
rotates with two angular momenta. Now solving the Euler-Lagrangian
equation for $t$ with appropriate integration constant we get the
equation of motion which is
\begin{eqnarray}
\dot{\psi^2} - \kappa^2\cosh^2\rho_0 &+& \frac{1}{2} (\mu^2_1 +
\mu^2_2) \sinh^2\rho_0 + m^2 \cosh^2\rho_0 \sin^2\psi \cr && \cr
&+& \nu^2_1 \sin^2\psi + \nu^2_2 \cos^2\psi = 0.\label{R}
\end{eqnarray}
The above equation (\ref{R}) gives us the potential energy
$V(\psi)$ which gives the oscillation number for the strings. If
we put $ \mu_i = \nu_i = 0$ in the above equation (\ref{R}), it
changes to
\begin{equation}
\dot{\psi^2} = \kappa^2 \cosh^2\rho_0 - m^2 \cosh^2\rho_0
\sin^2\psi ,\label{S}
\end{equation}
which is similar to the conformal gauge condition obtained in
\cite{Beccaria:2010zn}. With $\mu_i = \nu_i =0$, the conserved
quantities $ S_{i=1,2}$ and $J_{i=1,2}$ become zero. The energy
and oscillation number imply
\begin{eqnarray}
\mathcal E &=& \frac{E}{\sqrt{\lambda}} = \kappa \cosh\rho_0, \cr
&& \cr \mathcal N &=& \frac{N}{\sqrt{\lambda}} = \frac{1}{2\pi}
\oint ~d\psi ~\dot{\psi} =
\frac{2n}{\pi}\Big[(\frac{\gamma^2}{n^2}-1)\mathbb{K}
(\frac{\gamma^2}{n^2}) + \mathbb{E}(\frac{\gamma^2}{n^2}) \Big],
\label{T}
\end{eqnarray}
where $\mathbb{K}$ and $\mathbb{E}$ are the usual elliptical
functions, $\gamma = \kappa\cosh\rho_0 $ and $n = m\cosh\rho_0$ .
For small $\gamma$ and $\mathcal N$, we get the classical energy
as
\begin{equation}
 \mathcal E = \sqrt{2 n \mathcal N} + O \Big(
\mathcal N^{\frac{3}{2}}\Big), \label{Y}
\end{equation}
which is similar to the relation got in \cite{Beccaria:2010zn}. If
we consider $\rho_0 = 0 $, then we get back the exact energy for
the string oscillating in the $R \times S^2$. It will be interesting
to find out solutions of open string that oscillates in AdS$_5$ and
at the same time rotates along S$^5$.
\section{Conclusions}
In this paper, we have found two classes of giant magnon and spiky
string solutions in $AdS_5 \times S^5$ background with one spin
along $AdS_5$ and three angular momenta in $S^5$. We have
calculated the divergent energies and angular momenta and have
found out new solutions corresponding to giant magnon and spiky
strings. The dispersion relations we got in (\ref{20}) and
(\ref{23}) are the generalized solutions of \cite{Giardino:2011uc}
to include a spin along $AdS_5$. Similarly, by putting $S_1 = 0$
in the (\ref{30}) we get the relation obtained in
\cite{Kruczenski:2006pk}. Though, in \cite{Kruczenski:2006pk} all
the three angular momenta are infinite, in our case one of the
momenta $(J_1)$ is finite. The dispersion relation found in the
(\ref{31}) is similar to \cite{Ryang:2006yq}, where we combine the
two momenta to get the energy of a superposition of two magnon
bound states. The last dispersion relation in (\ref{33}) presents
a new class of spiky string solution. However, with a different
constraints of integration constants we can get a dispersion
relation same as obtained in (\ref{23}). There are further questions
that can be addressed. First of all it will be interesting to look
for solutions with divergent energy and multiple angular momenta in the
$\beta$ deformed $AdS_5 \times S^5$. However, the presence of various
background fluxes makes the study more difficult, nevertheless worth
exploring. According to AdS/CFT duality these states would correspond
to some gauge theory operators, though the exact nature of the operators
are not known, but it will most likely fall into the class of operators that
correspond to the known multispin solutions on the $AdS_5 \times S^5$.
It will be interesting to look at the operator dual of such
states in detail. For example the oscillating and rotating string solutions present in
the last section might correspond to certain highly excited sigma model operators.
These in turn must have something to do with the higher spin states and it will be
exciting to study the nature of these states and operators.
\vskip .2in
\noindent
{\bf Acknowledgements:} KLP would like to thank the Abdus Salam I.C.T.P, Trieste for hospitality under Associate
Scheme, where a part of this work was completed.


\begin{thebibliography}{99}
%\cite{Maldacena:1997re}
\bibitem{Maldacena:1997re}
  J.~M.~Maldacena,
  ``The Large N limit of superconformal field theories and supergravity,''
  Adv.\ Theor.\ Math.\ Phys.\  {\bf 2}, 231 (1998)
  [Int.\ J.\ Theor.\ Phys.\  {\bf 38}, 1113 (1999)]
  [hep-th/9711200].
  %%CITATION = HEP-TH/9711200;%%

%\cite{Gubser:1998bc}
\bibitem{Gubser:1998bc}
  S.~S.~Gubser, I.~R.~Klebanov and A.~M.~Polyakov,
  ``Gauge theory correlators from noncritical string theory,''
  Phys.\ Lett.\ B {\bf 428} (1998) 105
  [hep-th/9802109].
  %%CITATION = HEP-TH/9802109;%%

%\cite{Witten:1998qj}
\bibitem{Witten:1998qj}
  E.~Witten,
  ``Anti-de Sitter space and holography,''
  Adv.\ Theor.\ Math.\ Phys.\  {\bf 2} (1998) 253
  [hep-th/9802150].
  %%CITATION = HEP-TH/9802150;%%

  %\cite{Berenstein:2002jq}
\bibitem{Berenstein:2002jq}
  D.~E.~Berenstein, J.~M.~Maldacena and H.~S.~Nastase,
  ``Strings in flat space and pp waves from N=4 superYang-Mills,''
  JHEP {\bf 0204}, 013 (2002)
  [hep-th/0202021].
  %%CITATION = HEP-TH/0202021;%%

%\cite{Gubser:2002tv}
\bibitem{Gubser:2002tv}
  S.~S.~Gubser, I.~R.~Klebanov and A.~M.~Polyakov,
  ``A Semiclassical limit of the gauge / string correspondence,''
  Nucl.\ Phys.\ B {\bf 636}, 99 (2002)
  [hep-th/0204051].
  %%CITATION = HEP-TH/0204051;%%

%\cite{Minahan:2002ve}
\bibitem{Minahan:2002ve}
  J.~A.~Minahan and K.~Zarembo,
  ``The Bethe ansatz for N=4 superYang-Mills,''
  JHEP {\bf 0303}, 013 (2003)
  [hep-th/0212208].
  %%CITATION = HEP-TH/0212208;%%


%\cite{Beisert:2003yb}
\bibitem{Beisert:2003yb}
  N.~Beisert and M.~Staudacher,
  ``The N=4 SYM integrable super spin chain,''
  Nucl.\ Phys.\ B {\bf 670}, 439 (2003)
  [hep-th/0307042].
  %%CITATION = HEP-TH/0307042;%%

%\cite{Bena:2003wd}
\bibitem{Bena:2003wd}
  I.~Bena, J.~Polchinski and R.~Roiban,
  ``Hidden symmetries of the AdS(5) x S**5 superstring,''
  Phys.\ Rev.\ D {\bf 69}, 046002 (2004)
  [hep-th/0305116].
  %%CITATION = HEP-TH/0305116;%%



%\cite{Kazakov:2004qf}
\bibitem{Kazakov:2004qf}
  V.~A.~Kazakov, A.~Marshakov, J.~A.~Minahan and K.~Zarembo,
  ``Classical/quantum integrability in AdS/CFT,''
  JHEP {\bf 0405}, 024 (2004)
  [hep-th/0402207].
  %%CITATION = HEP-TH/0402207;%%

%\cite{Zarembo:2004hp}
\bibitem{Zarembo:2004hp}
  K.~Zarembo,
  ``Semiclassical Bethe Ansatz and AdS/CFT,''
  Comptes Rendus Physique {\bf 5}, 1081 (2004)
  [Fortsch.\ Phys.\  {\bf 53}, 647 (2005)]
  [hep-th/0411191].
  %%CITATION = HEP-TH/0411191;%%

%\cite{Frolov:2002av}
\bibitem{Frolov:2002av}
  S.~Frolov and A.~A.~Tseytlin,
  ``Semiclassical quantization of rotating superstring in AdS(5) x S**5,''
  JHEP {\bf 0206}, 007 (2002)
  [hep-th/0204226].
  %%CITATION = HEP-TH/0204226;%%

%\cite{Frolov:2003qc}
\bibitem{Frolov:2003qc}
  S.~Frolov and A.~A.~Tseytlin,
  ``Multispin string solutions in AdS(5) x S**5,''
  Nucl.\ Phys.\ B {\bf 668}, 77 (2003)
  [hep-th/0304255].
  %%CITATION = HEP-TH/0304255;%%

%\cite{Frolov:2003tu}
\bibitem{Frolov:2003tu}
  S.~Frolov and A.~A.~Tseytlin,
  ``Quantizing three spin string solution in AdS(5) x S**5,''
  JHEP {\bf 0307}, 016 (2003)
  [hep-th/0306130].
  %%CITATION = HEP-TH/0306130;%%



%\cite{Tseytlin:2003ii}
\bibitem{Tseytlin:2003ii}
  A.~A.~Tseytlin,
  ``Spinning strings and AdS / CFT duality,''
  In *Shifman, M. (ed.) et al.: From fields to strings, vol. 2* 1648-1707
  [hep-th/0311139].
  %%CITATION = HEP-TH/0311139;%%

%\cite{Beisert:2004ry}
\bibitem{Beisert:2004ry}
  N.~Beisert,
  ``The Dilatation operator of N=4 super Yang-Mills theory and integrability,''
  Phys.\ Rept.\  {\bf 405}, 1 (2005)
  [hep-th/0407277].
  %%CITATION = HEP-TH/0407277;%%

%\cite{Hofman:2006xt}
\bibitem{Hofman:2006xt}
  D.~M.~Hofman and J.~M.~Maldacena,
  ``Giant Magnons,''
  J.\ Phys.\ A A {\bf 39}, 13095 (2006)
  [hep-th/0604135].
  %%CITATION = HEP-TH/0604135;%%


%\cite{Tseytlin:2004xa}
\bibitem{Tseytlin:2004xa}
  A.~A.~Tseytlin,
  ``Semiclassical strings and AdS/CFT,''
  [hep-th/0409296].
  %%CITATION = HEP-TH/0409296;%%

%\cite{Arutyunov:2006gs}
\bibitem{Arutyunov:2006gs}
  G.~Arutyunov, S.~Frolov and M.~Zamaklar,
  ``Finite-size Effects from Giant Magnons,''
  Nucl.\ Phys.\ B {\bf 778}, 1 (2007)
  [hep-th/0606126].
  %%CITATION = HEP-TH/0606126;%%

%\cite{Minahan:2006bd}
\bibitem{Minahan:2006bd}
  J.~A.~Minahan, A.~Tirziu and A.~A.~Tseytlin,
  ``Infinite spin limit of semiclassical string states,''
  JHEP {\bf 0608}, 049 (2006)
  [hep-th/0606145].
  %%CITATION = HEP-TH/0606145;%%

%\cite{Dorey:2006dq}
\bibitem{Dorey:2006dq}
  N.~Dorey,
  ``Magnon Bound States and the AdS/CFT Correspondence,''
  J.\ Phys.\ A A {\bf 39}, 13119 (2006)
  [hep-th/0604175].
  %%CITATION = HEP-TH/0604175;%%

%\cite{Chen:2006gea}
\bibitem{Chen:2006gea}
  H.~-Y.~Chen, N.~Dorey and K.~Okamura,
  ``Dyonic giant magnons,''
  JHEP {\bf 0609}, 024 (2006)
  [hep-th/0605155].
  %%CITATION = HEP-TH/0605155;%%

%\cite{Kruczenski:2006pk}
\bibitem{Kruczenski:2006pk}
  M.~Kruczenski, J.~Russo and A.~A.~Tseytlin,
  ``Spiky strings and giant magnons on S**5,''
  JHEP {\bf 0610}, 002 (2006)
  [hep-th/0607044].
  %%CITATION = HEP-TH/0607044;%%

%\cite{Ryang:2006yq}
\bibitem{Ryang:2006yq}
  S.~Ryang,
  ``Three-spin giant magnons in AdS(5) x S**5,''
  JHEP {\bf 0612}, 043 (2006)
  [hep-th/0610037].
  %%CITATION = HEP-TH/0610037;%

%\cite{Kruczenski:2004wg}
\bibitem{Kruczenski:2004wg}
  M.~Kruczenski,
  ``Spiky strings and single trace operators in gauge theories,''
  JHEP {\bf 0508}, 014 (2005)
  [hep-th/0410226].
  %%CITATION = HEP-TH/0410226;%%

%\cite{Ishizeki:2007we}
\bibitem{Ishizeki:2007we}
  R.~Ishizeki and M.~Kruczenski,
  ``Single spike solutions for strings on S**2 and S**3,''
  Phys.\ Rev.\ D {\bf 76}, 126006 (2007)
  [arXiv:0705.2429 [hep-th]].
  %%CITATION = ARXIV:0705.2429;%%

%\cite{Lunin:2005jy}
\bibitem{Lunin:2005jy}
 O.~Lunin and J.~M.~Maldacena,
  ``Deforming field theories with U(1) x U(1) global symmetry and their gravity duals,''
  JHEP {\bf 0505}, 033 (2005)
  [hep-th/0502086].
  %%CITATION = HEP-TH/0502086;%%

%\cite{Bobev:2005cz}
\bibitem{Bobev:2005cz}
  N.~P.~Bobev, H.~Dimov, R.~C.~Rashkov,
  ``Semiclassical strings in Lunin-Maldacena background,''
  [hep-th/0506063].

%\cite{Chu:2006ae}
\bibitem{Chu:2006ae}
  C.~-S.~Chu, G.~Georgiou and V.~V.~Khoze,
  ``Magnons, classical strings and beta-deformations,''
  JHEP {\bf 0611}, 093 (2006)
  [hep-th/0606220].
  %%CITATION = HEP-TH/0606220;%%

%\cite{Bobev:2006fg}
\bibitem{Bobev:2006fg}
  N.~P.~Bobev and R.~C.~Rashkov,
  ``Multispin Giant Magnons,''
  Phys.\ Rev.\ D {\bf 74}, 046011 (2006)
  [hep-th/0607018].
  %%CITATION = HEP-TH/0607018;%%

%\cite{Kluson:2007qu}
\bibitem{Kluson:2007qu}
  J.~Kluson, R.~R.~Nayak, K.~L.~Panigrahi,
  ``Giant Magnon in NS5-brane Background,''
  JHEP {\bf 0704}, 099 (2007).
  [hep-th/0703244].


%\cite{Lee:2008sk}
\bibitem{Lee:2008sk}
  B.~-H.~Lee, R.~R.~Nayak, K.~L.~Panigrahi, C.~Park,
  ``On the giant magnon and spike solutions for strings on AdS(3) x S**3,''
  JHEP {\bf 0806}, 065 (2008).
  [arXiv:0804.2923 [hep-th]].

%\cite{David:2008yk}
\bibitem{David:2008yk}
  J.~R.~David, B.~Sahoo,
  ``Giant magnons in the D1-D5 system,''
  JHEP {\bf 0807}, 033 (2008).
  [arXiv:0804.3267 [hep-th]].

%\cite{Grignani:2008is}
\bibitem{Grignani:2008is}
  G.~Grignani, T.~Harmark and M.~Orselli,
  ``The SU(2) x SU(2) sector in the string dual of N=6 superconformal Chern-Simons theory,''
  Nucl.\ Phys.\ B {\bf 810}, 115 (2009)
  [arXiv:0806.4959 [hep-th]].
  %%CITATION = ARXIV:0806.4959;%%

%\cite{Lee:2008ui}
\bibitem{Lee:2008ui}
  B.~-H.~Lee, K.~L.~Panigrahi, C.~Park,
  ``Spiky Strings on AdS(4) x CP**3,''
  JHEP {\bf 0811}, 066 (2008).
  [arXiv:0807.2559 [hep-th]].

%\cite{Ryang:2008rc}
\bibitem{Ryang:2008rc}
  S.~Ryang,
  ``Giant Magnon and Spike Solutions with Two Spins in AdS(4) x CP**3,''
  JHEP {\bf 0811}, 084 (2008).
  [arXiv:0809.5106 [hep-th]].

%\cite{Benvenuti:2008bd}
\bibitem{Benvenuti:2008bd}
  S.~Benvenuti and E.~Tonni,
  ``Giant magnons and spiky strings on the conifold,''
  JHEP {\bf 0902}, 041 (2009)
  [arXiv:0811.0145 [hep-th]].
  %%CITATION = ARXIV:0811.0145;%%

%\cite{Abbott:2008qd}
\bibitem{Abbott:2008qd}
  M.~C.~Abbott, I.~Aniceto,
  ``Giant Magnons in AdS(4) x CP**3: Embeddings, Charges and a Hamiltonian,''
  JHEP {\bf 0904}, 136 (2009).
  [arXiv:0811.2423 [hep-th]].

%\cite{Biswas:2011wu}
\bibitem{Biswas:2011wu}
 S.~Biswas and K.~L.~Panigrahi,
  ``Spiky Strings on NS5-branes,''
  Phys.\ Lett.\ B {\bf 701}, 481 (2011)
  [arXiv:1103.6153 [hep-th]],
  %%CITATION = ARXIV:1103.6153;%%

  %\cite{Biswas:2012wu}
\bibitem{Biswas:2012wu}
  S.~Biswas and K.~L.~Panigrahi,
  ``Spiky Strings on I-brane,''
  arXiv:1206.2539 [hep-th].
  %%CITATION = ARXIV:1206.2539;%%


%\cite{Giardino:2011dz}
\bibitem{Giardino:2011dz}
  S.~Giardino and H.~L.~Carrion,
  ``Classical strings in AdS(4) x CP(3) with three angular momenta,''
  JHEP {\bf 1108}, 057 (2011)
  [arXiv:1106.5684 [hep-th]].
  %%CITATION = ARXIV:1106.5684;%%

%\cite{Panigrahi:2011be}
\bibitem{Panigrahi:2011be}
  K.~L.~Panigrahi, P.~M.~Pradhan and P.~K.~Swain,
  ``Rotating Strings in AdS(4) x CP(3) with B(NS) holonomy,''
  JHEP {\bf 1201}, 113 (2012)
  [arXiv:1109.2458 [hep-th]].
  %%CITATION = ARXIV:1109.2458;%%

%\cite{Giardino:2011uc}
\bibitem{Giardino:2011uc}
  S.~Giardino,
  ``Divergent energy strings in $AdS_5\times S^5$ with three angular momenta,''
  JHEP {\bf 1112}, 022 (2011)
  [arXiv:1110.3682 [hep-th]].
  %%CITATION = ARXIV:1110.3682;%%

%\cite{Panigrahi:2012bm}
\bibitem{Panigrahi:2012bm}
  K.~L.~Panigrahi, P.~M.~Pradhan and P.~K.~Swain,
  ``Three Spin Spiky Strings in $\beta$-deformed Background,''
  JHEP {\bf 1206}, 057 (2012)
  [arXiv:1203.3057 [hep-th]].
  %%CITATION = ARXIV:1203.3057;%%

%\cite{Minahan:2002rc}
\bibitem{Minahan:2002rc}
  J.~A.~Minahan,
  ``Circular semiclassical string solutions on AdS(5) x S(5),''
  Nucl.\ Phys.\ B {\bf 648}, 203 (2003)
  [hep-th/0209047].
  %%CITATION = HEP-TH/0209047;%%

%\cite{Beccaria:2010zn}
\bibitem{Beccaria:2010zn}
  M.~Beccaria, G.~V.~Dunne, G.~Macorini, A.~Tirziu and A.~A.~Tseytlin,
  ``Exact computation of one-loop correction to energy of pulsating strings in $AdS_5 x S^5$,''
  J.\ Phys.\ A A {\bf 44}, 015404 (2011)
  [arXiv:1009.2318 [hep-th]].
  %%CITATION = ARXIV:1009.2318;%%

%\cite{Park:2005kt}
\bibitem{Park:2005kt}
  I.~Y.~Park, A.~Tirziu and A.~A.~Tseytlin,
  ``Semiclassical circular strings in AdS(5) and 'long' gauge field strength operators,''
  Phys.\ Rev.\ D {\bf 71}, 126008 (2005)
  [hep-th/0505130].
  %%CITATION = HEP-TH/0505130;%%
\end{thebibliography}
\end{document}